\documentclass{article}
\usepackage{spconf,amsmath,graphicx,epsfig}
\usepackage{verbatim}
\usepackage{fancyhdr}
\usepackage{amsfonts,todonotes}
\usepackage[algoruled,boxed,lined]{algorithm2e}
\usepackage{balance}
\usepackage{hyperref}
\usepackage{enumitem}
\makeatletter
\def\BState{\State\hskip-\ALG@thistlm}

\newcommand\blfootnote[1]{%
  \begingroup
  \renewcommand\thefootnote{}\footnote{#1}%
  \addtocounter{footnote}{-1}%
  \endgroup
}

\SetKwProg{Fn}{Function}{}{}

\usepackage{pifont}
\newcommand{\cmark}{\ding{51}}%
\newcommand{\xmark}{\ding{55}}%

\fancypagestyle{firstpage}{%
  \lfoot{\small{\textit{To appear in the Proceedings of ICASSP, 2019, Brighton, UK}}}
}

\title{LEARNING AFFECTIVE CORRESPONDENCE BETWEEN MUSIC AND IMAGE}
\name{Gaurav Verma$^1$, Eeshan Gunesh Dhekane$^2$, Tanaya Guha$^3$}
\address{ 
$^{1}$Adobe Research, India\\
$^{2}$Mila, Universit\'e de Montr\'eal, Canada\\
$^{3}$University of Warwick, UK\\
}

\begin{document}
\ninept
\maketitle
\thispagestyle{firstpage}
\begin{abstract}
We introduce the problem of learning \emph{affective correspondence} between audio (music) and visual data (images). For this task, a music clip and an image are considered \emph{similar} (having true correspondence) if they have similar emotion content. In order to estimate this crossmodal, emotion-centric similarity, we propose a deep neural network architecture that learns to project the data from the two modalities to a common representation space, and performs a binary classification task of predicting the affective correspondence (true or false). To facilitate the current study, we construct a large scale database containing more than $3,500$ music clips and $85,000$ images with three emotion classes (positive, neutral, negative). The proposed approach achieves  $61.67\%$ accuracy for the affective correspondence prediction task on this database, outperforming two relevant and competitive baselines. We also demonstrate that our network learns modality-specific representations of emotion (without explicitly being trained with emotion labels), which are useful for emotion recognition in individual modalities.
\end{abstract}
\section{Introduction}
\vspace{-1.0mm}
\label{sec:intro}
\blfootnote{A part of this work was done when all the authors were with Indian Institute of Technology (IIT) Kanpur, India.}
Automatic analysis of emotional content in data is important for developing human-centric intelligent systems. In general, emotion recognition is a challenging task due to the huge variability and subjectivity involved in the expression and perception of human emotion. In recent years, significant progress has been made towards the recognition and analysis of emotion in individual modalities, such as in images and videos \cite{joshi2011aesthetics,you2015robust,kahou2013combining}, and in speech and music \cite{schuller2003hidden,yang2011ranking}. Since human emotion is inherently multimodal, research efforts that combine information from multiple modalities  are also on the rise~\cite{socher2013zero,owens2016ambient,goyal2016multimodal,cao2016deep,hong2017deep}. 
\par 
\begin{figure}[tb]
\centering
\includegraphics[width=0.9\columnwidth]{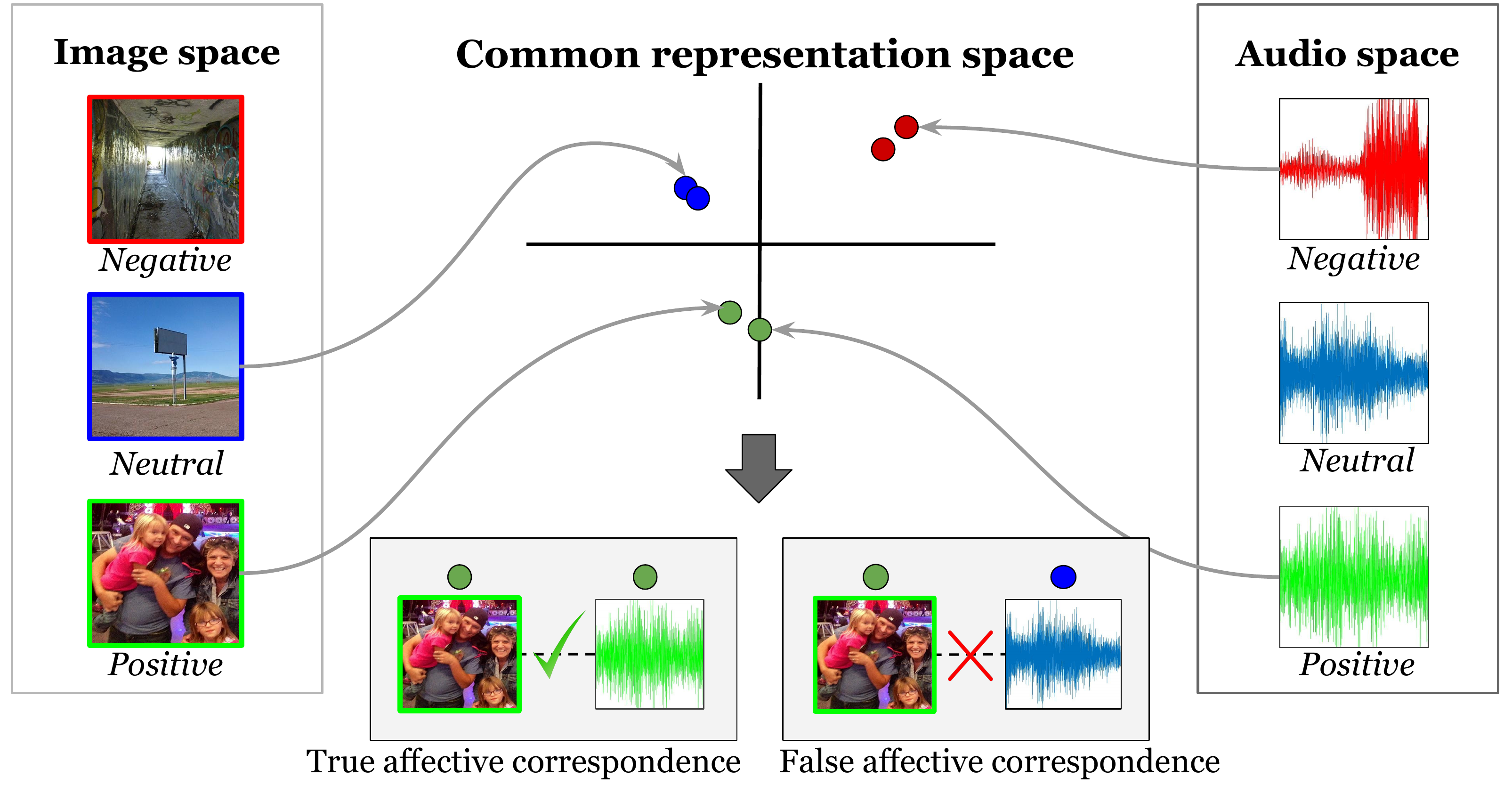}
\caption{The basic idea behind our affective correspondence prediction framework. }
\label{fig:teaser}
\end{figure}
The majority of existing multimodal systems focus on achieving better emotion recognition accuracy by fusing information from different modalities. Relatively less effort has been put towards understanding the emotion-centric relationship between modalities, and how the emotional content is shared across multiple modalities.  In general, crossmodal work involving audio and visual data has been quite limited. Very recently, researchers have started exploring the crossmodal relationship between audio and visual data for different applications. Owens et al.~\cite{owens2018audio} use convolutional neural networks (CNNs) to predict, in a self-supervised way, if a given pair of audio and video clip is temporally aligned or not. The learned representations are subsequently used to perform sound source localization, and audio-visual action recognition. In a task of crossmodal biometric matching, Nagrani et al. \cite{nagrani2018seeing} propose to match a given voice sample against two or more faces. The work closest to our task is the work of Arandjelovi\'c et al. \cite{arandjelovic2017look}, where the task of audio-visual correspondence learning was introduced. A deep neural network model that comprises visual and audio subnetworks was trained to learn semantic correspondence between audio and visual data. In another related work, Arandjelovi\'c et al. \cite{arandjelovic2017objects} propose a network that can localize the objects in an image corresponding to a sound input. However, none of the above work has studied the relationship across modalities from the perspective of emotion. 
\par

In this paper, we introduce the task of learning \emph{affective correspondence} between audio (music) and visual data (images). We consider a music clip and an image to be \emph{similar} (true correspondence) if they are known to evoke the same broad category of emotion, and to be dissimilar otherwise. Our objective is to build a model that can identify whether or not a music-image pair contains similar information in terms of emotion. The basic idea is to project the image and music data to a common representation space where their emotional content can be compared (see Fig.~\ref{fig:teaser}). An effective solution to this problem will be useful for crossmodal retrieval and emotion-aware recommendation systems. In order to estimate the emotion-centric similarity between music-image pairs, we propose a deep neural network architecture that learns to compare the emotional content present in the two modalities without explicitly requiring emotion labels. Fig.~\ref{fig:architecture} presents an overview of the proposed network architecture. Our network consists of two subnetworks pertaining to the visual and music modalities that project the two disparate modalities to a common representation space. These representations are used to learn the crossmodal correspondence by training the network to perform a binary classification task of predicting the affective correspondence (true or false) between music-image pairs. To facilitate the current study, we have constructed a large scale database comprising music and images with three emotion labels - positive, neutral and negative. We evaluate the performance of our approach on this database primarily for the affective correspondence task, and compare it with relevant and competitive baselines. In addition, we demonstrate that the intermediate modality-specific representations learned by our network are also useful for emotion recognition in music or images. 
\section{Cross-Modal Database Creation}
\label{sec:database}
\begin{table}[tb]
	\caption{Regrouping of image labels from the Image Emotion database \cite{HailinJin} to our IMAC database.}
    \vspace{2mm}
    \scriptsize
    \centering
    \resizebox{0.75\columnwidth}{!}{%
    \begin{tabular}{c c}\hline
         \textbf{Original label} &  \textbf{Emotion class} \\\hline
         Awe, amusement, excitement & Positive  \\
         Contentment  & Neutral \\
         Fear, disgust, anger, sadness & Negative \\ \hline
    \end{tabular}%
    }
    \label{tab:IED_mapping}
    \vspace{-3mm}
\end{table}
To facilitate the study of crossmodal emotion analysis, and owing to the lack of relevant databases, we constructed a large scale database, which we call the \emph{Image-Music Affective Correspondence} (IMAC) database\footnote{Our IMAC database is available at \url{https://gaurav22verma.github.io/IMAC_Dataset.html}}. It consists of more than $85,000$ 
images and $3,812$ songs ($\sim 270$ hours of audio). Each data sample is labeled with one of the three emotions: positive, neutral and negative. The IMAC database is constructed by combining an existing image emotion database \cite{HailinJin} with a new music emotion database curated by the authors of this paper. Below, we describe the construction of the database in detail.\\

\vspace{-3mm}\noindent\textbf{Image data collection and labeling:}
For image data, we use the Image Emotion database curated by You et al. \cite{HailinJin}. This database contains over $85,000$ natural images labeled with one of the following emotions: amusement, anger, awe, contentment, disgust, excitement, fear, and sadness. Since the images have no manual annotation, they are considered to be weakly labeled. Of all these images, around $23,000$ images were labeled by humans through Amazon Mechanical Turk. For simplicity and higher interpretability, we regroup all the images into three broad emotion classes (\emph{positive, neutral} and \emph{negative}) using their original labels as shown in Table \ref{tab:IED_mapping}.\\

\vspace{-3mm}\noindent\textbf{Music data collection and labeling:} We created a Music Emotion database by collecting $3,812$ songs from YouTube, and labeling them using semi-automated techniques. These songs were chosen from the Million Song database \cite{bertin2011million} which provides various meta information for one million contemporary music tracks. Since manual labeling of such a large corpus is difficult and time consuming, we exploited the user tags made available for the soundtracks.

The first step towards labeling was to screen the tags that are relevant to our task, i.e., the tags related to emotion. We automatically selected the tags that contain any of the following strings: \textit{sad}, \textit{pain}, \textit{soothing}, \textit{relax}, \textit{calm}, \textit{happy}, \textit{joyous}, \textit{energetic}. After this step, the tags with irrelevant and ambiguous information (e.g., 'happysad') were manually removed. For higher interpretability, we grouped the tags into three broad categories: (i) \emph{positive}: includes the strings `happy', `joyous', and `energetic'; (ii) \emph{neutral}: includes the strings `soothing', `relax' and `calm'; (iii) \emph{negative}: includes the strings `sad' and `pain'. After this processing, we are left with $2971$ tags ($952$ positive, $752$ neutral, and $1267$ negative tags) in total. All the songs from the Million Song database associated with one or more of these tags were collected automatically by querying the YouTube API.
Finally, we shortlisted a total of $3,812$ songs, which collectively amount to about $270$ hours of audio data. Table \ref{tab:example_tags} presents a list of representative tags along with the number of times they appear in our database. 
\begin{table}[tb]
\caption{Examples of relevant user tags with their frequency of occurrence in our Music Emotion database.}
\vspace{2mm}
    \centering
   \resizebox{0.97\columnwidth}{!}{%
    \begin{tabular}{c c c}
    \hline
         \textbf{Emotion class} & \textbf{User tag} & \textbf{Frequency}\\
         \hline
         Positive & \emph{this will always make me happy} & 1\\
         & \emph{so energetic} & 26\\
         & \emph{makes me energetic and wanna dance} & 3\\
         & \emph{joyous} & 103\\ \hline
         Neutral & \emph{soothing for the ear to hear} & 18\\
         & \emph{cool and relaxing music} & 14\\
         & \emph{calmness} & 43\\ \hline
          Negative & \emph{sad} & 873  \\
         & \emph{makes me sad} & 89 \\
         & \emph{for the painfully alone} & 34\\\hline
    \end{tabular}%
    }
    \label{tab:example_tags}
\end{table}
Due to the subjectivity involved in emotion perception, it is possible for a single song to have more than one type of tags (e.g., positive and neutral) associated with it. In such cases, we select the most dominant tag as the label of the song. In case of a tie, preference is given to the more positive tag.\\

\vspace{-2mm}\noindent\textbf{Affective correspondence labeling:} 
Our IMAC database combines the data from the Image Emotion database \cite{HailinJin}, and the newly curated Music Emotion database. As described above, each sample in both the databases is now labeled with one of the three broad emotion classes: \emph{positive}, \emph{neutral} and \emph{negative} (see Table \ref{tab:imac}). We consider an image-music pair to have true \emph{affective correspondence} if and only if both belong to the same broad class.
\begin{table}[tb]
\caption{Details of the IMAC database.}
\label{tab:imac}
    \vspace{2mm}
    \centering
    \begin{tabular}{c c c}
    \hline
    \bf Emotion class & \bf Image samples & \bf Music samples\\
    \hline
    \emph{positive} & 32,868	& 1,342 \\
    \emph{neutral} & 10,796	& 1,511\\
    \emph{negative} & 45,731 & 959 \\
    \hline
    \end{tabular}
\end{table}
\section{Affective Correspondence Learning}
\label{sec:approach}

\begin{figure}[!t]
\centering
\includegraphics[width=0.73\columnwidth]{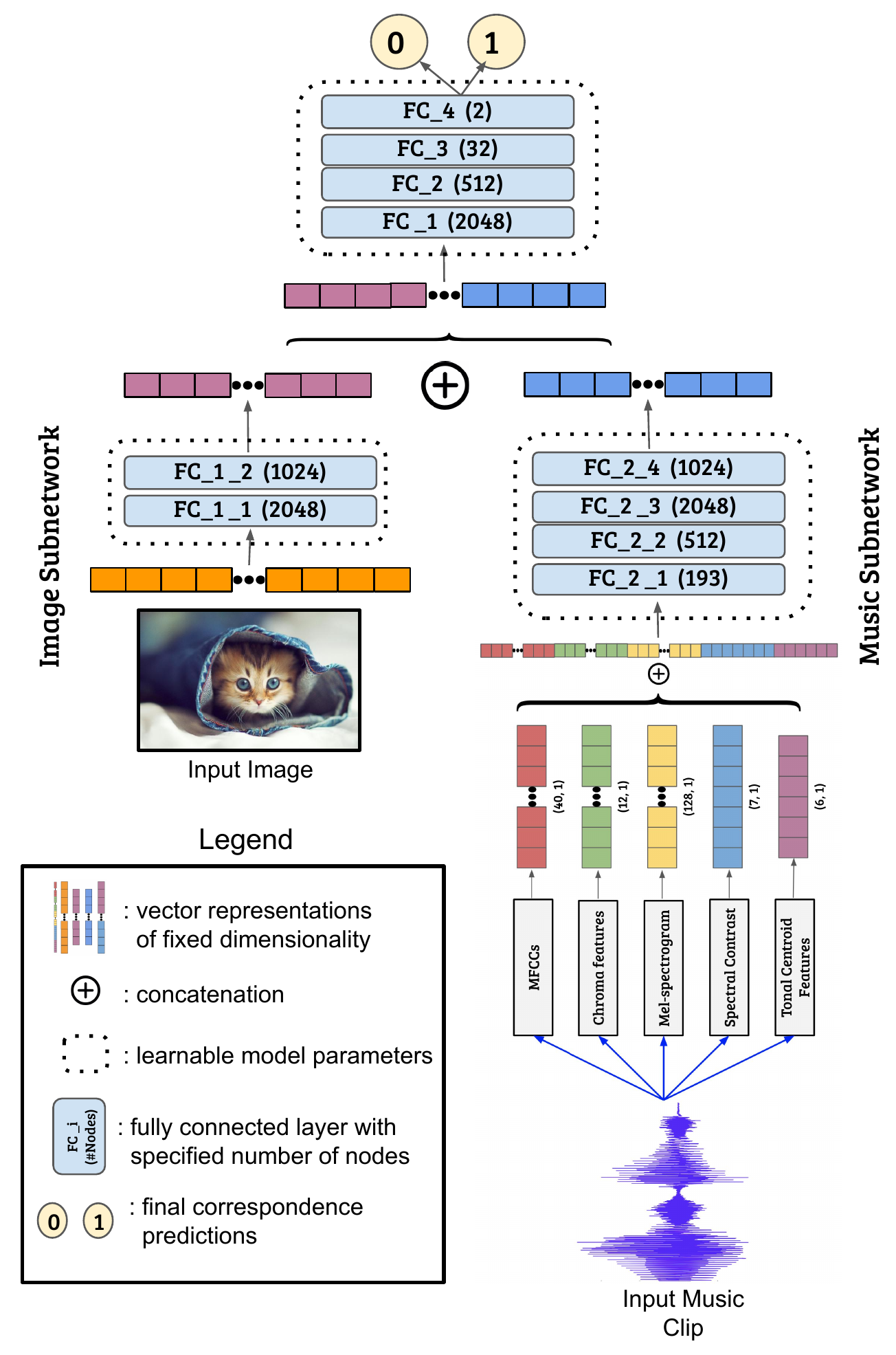}
\caption{Proposed architecture for affective correspondence learning} 
\label{fig:architecture}
\end{figure}

Our objective is to build a network that can identify if an image-music pair has similar emotional content or not. Let $\mathcal{C}$ be the set of all emotion labels, and $\mathcal{X}$ and $\mathcal{Y}$ denote the sets of all image and music samples. Each image $\mathbf{x}\in\mathcal{X}$ and music sample $\mathbf{y}\in\mathcal{Y}$ has a unique emotion label $c(\mathbf{x})\in\mathcal{C}$, $c(\mathbf{y})\in\mathcal{C}$. We intend to learn a mapping $f : \mathcal{X}\times \mathcal{Y}\rightarrow \{0, 1\}$ such that for each pair $(\mathbf{x}, \mathbf{y})$, $f(\mathbf{x}, \mathbf{y}) = 1$ (true correspondence) if $c(\mathbf{x}) =c(\mathbf{y})$, and $f(\mathbf{x}, \mathbf{y}) = 0$ (false correspondence) otherwise. With two significantly different modalities in hand, we propose to first project the data from each modality to a common space $\mathcal{V}$, and then perform a binary classification on the learned representations. We define two mappings corresponding to the two modalities: $f_x:\mathcal{X}\rightarrow\mathcal{V}$ and $f_y:\mathcal{Y}\rightarrow\mathcal{V}$. Consider a binary classifier $g:\mathcal{V}\oplus\mathcal{V}\rightarrow\{0, 1\}$ that operates on the concatenated multimodal representation in the common space $\mathcal{V}\oplus\mathcal{V}$. Then, the original task of learning $f$ can be rewritten as $f = g\circ\left(f_x\oplus f_y\right)$. The proposed \emph{affective correspondence prediction} network (ACPnet) in Fig.~\ref{fig:architecture} models $f_x$, $f_y$  and $g$. The network is trained end-to-end using the IMAC database described in Section \ref{sec:database}.
The proposed ACP-Net has two subnetworks corresponding to the two modalities, which are connected to several fusion layers that combine the representations from images and music. Below, we describe each of these parts in detail. \\

\vspace{-2mm}\noindent\textbf{Image subnetwork:}
The image subnetwork relies on the strength of transfer learning, where the features learned for one task is used as an initialization for another task. In order to extract features from the images in our database, we use the \emph{Inception-V3} network \cite{szegedy2016rethinking} pretrained on the Imagenet database for the task of object classification \cite{russakovsky2015imagenet}. We hypothesize that the semantic features learned by the pretrained Inception network will be useful for emotion-related tasks as well. The image features are obtained by feedforwarding the images in our images through the pretrained network and using the representation from the last fully connected (FC) layer before the classification layer. These features are input to two FC layers yielding a final $1024$-dimensional representation (see Fig.~\ref{fig:architecture}).\\

\vspace{-2mm}\noindent\textbf{Music subnetwork:} For the music samples, we extract a set of acoustic features that are widely used for emotion recognition in audio \cite{west2004features, lu2006automatic, scherer1996adding}. We obtain the following features from the music clips: melfrequency cepstral coefficients (MFCCs), chroma features, spectral contrast features \cite{jiang2002music}, tonal centroid features \cite{harte2006detecting}, and additional features from the mel-spectrogram. In total, this yields a $193$-dimensional feature vector per music sample. This feature vector is input to $4$ FC layers yielding a final $1024$-dimensional representation (see Fig.~\ref{fig:architecture}).\\

\vspace{-2mm}\noindent\textbf{Fusion layers:} The fusion layers combine the image and music subnetworks which model the transform functions $f_x$ and $f_y$ respectively. Each subnetwork generates an embedding of its respective modality which is in the common representation space $\mathcal{V}\in\mathbb{R}^{1024}$. We concatenate the embeddings to get a multimodal representation $f_x(\mathbf{x}) \oplus f_y(\mathbf{y})$. The concatenated feature is passed through $4$ FC fusion layers (see Fig.~\ref{fig:architecture}) which produces a binary label to reflect the affective correspondence between the two modalities.
\section{Performance Evaluation}
\label{sec: expertiments}
\begin{figure*}[!t]
    \centering
    \begin{minipage}{1.0\textwidth}
        \centering
        \includegraphics[width=0.84\textwidth]{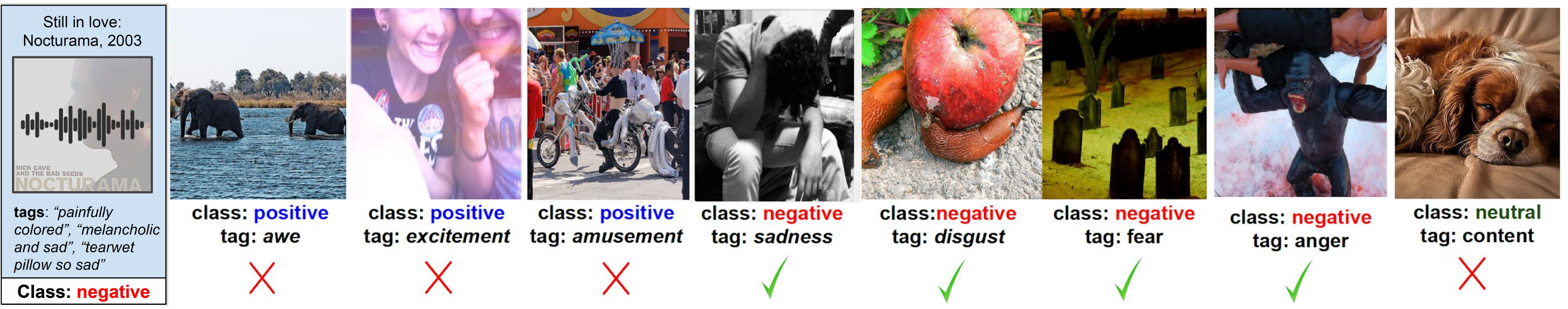}
    \end{minipage}\hfill
    \begin{minipage}{1.0\textwidth}
        \centering
        \includegraphics[width=0.84\textwidth]{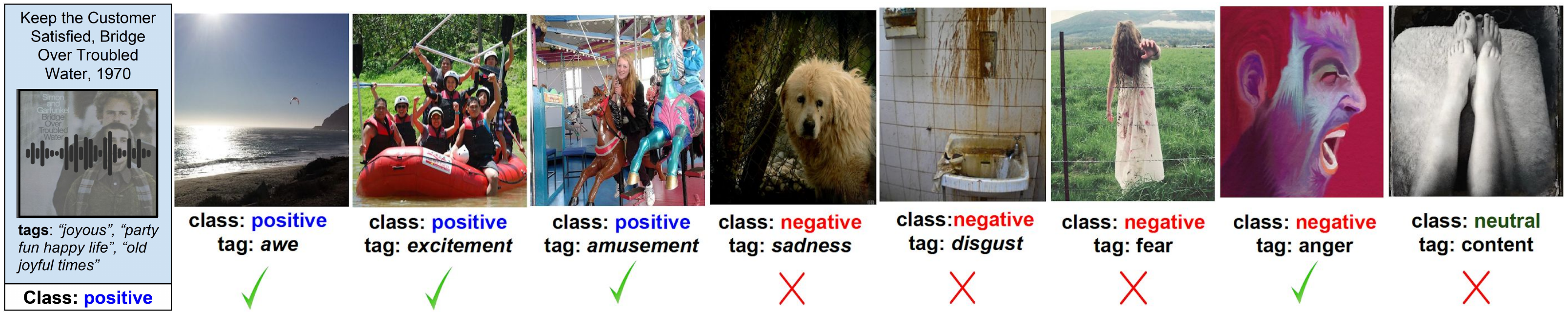} 
    \end{minipage}\hfill
    \begin{minipage}{1.0\textwidth}
        \centering
        \includegraphics[width=0.84\textwidth]{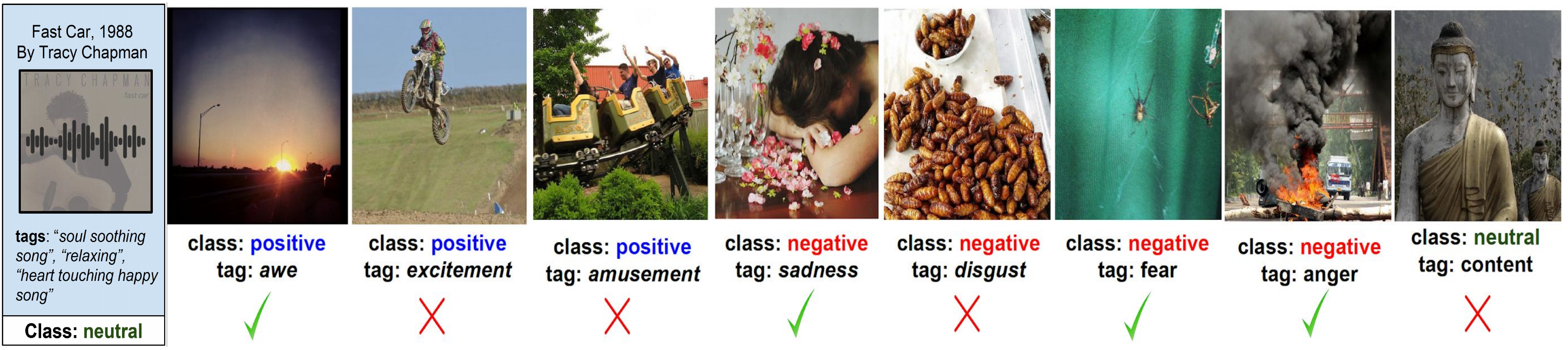}
    \end{minipage}
\caption{Affective correspondence prediction results achieved by our method. The \cmark and \xmark$\text{ }$ marks denote true and false correspondence.}
\label{fig:qual-eval}
\end{figure*}
In this section, we provide the details of our experimental settings, and present results for affective correspondence as well as emotion recognition in images and music. \\

\vspace{-2mm}\noindent\textbf{Database, experimental setup:} The proposed ACP network is trained and evaluated on the IMAC database we constructed (see Section \ref{sec:database}). The images and audio samples in the database are randomly distributed into training, validation and test sets with ratios 70:10:20. The model parameters are updated to minimize the cross-entropy loss between the predicted and true correspondence labels. We use Adam optimizer \cite{kingma2014adam} with a learning rate initialized at $10^{-4}$, and dropout for regularization with the probability of dropping a node as $0.4$. We train the models for a maximum of $50$ epochs with early stopping based on the accuracy achieved on the validation set.

For homogeneity, each original music clip was partitioned into non-overlapping segments of duration $60$ seconds each. Each $60$s long segment is treated as an independent data sample with the same label. The local acoustic features are extracted from each music segment using a window size of $10$s with $5$s overlap, and the mean of these local acoustic features is used to obtain a global feature representation. The window size is set to be longer than that typically used in audio processing. This is motivated by the observation that emotion is a smoothly varying function, and requires a longer context to be captured meaningfully \cite{gupta2014multimodal}. The final $193$-dimensional music feature vector is composed of $40$ MFCCs, $12$ chroma features, $7$ spectral contrast features, $6$ tonal centroid features, and $128$ features obtained from the mel spectrogram. Each image is also resized to a predefined size of $[224\times224\times3]$ before they are passed to the Inception-v3 network. The proposed ACP-Net is trained and validated on the IMAC database.\\

\vspace{-2mm}\noindent\textbf{Results on affective correspondence:} We compare the performance of our best-trained ACP-Net model with (i) the $L^3$-Net \cite{arandjelovic2017look}, and (ii) a variant of the ACP-Net that uses spectrograms as input to the music subnetwork.  We implement the $L^3$-Net as described in the original paper, and train it on our database. $L^3$-Net uses raw images and log-spectrograms as input to the image and audio subnetworks. The spectrograms were created by performing short-time Fourier transform with an window size of $100$ms and a hop length of $25$ms. All the networks are trained and validated on the same data, and the results on the test set are reported.
\begin{table}[tb]
\caption{Affective correspondence prediction results on the IMAC database}
\label{tab:affect-correspondence}
    \vspace{1mm}
    \centering
    \begin{tabular}{l c }
    \hline
    \bf Model & {\bf Accuracy} (in \%) \\
    \hline
    $L^3$-Net \cite{arandjelovic2017look} & 57.17\\
    Spectrogram-ACP & 55.84 \\
    {\bf ACP-Net} (proposed) & \textbf{61.67} \\
    \hline
    \end{tabular}
\end{table}
Table \ref{tab:affect-correspondence} compares the performance of the ACP-Net with above methods in terms of correspondence prediction accuracy. The results show that the proposed ACP-Net outperforms the $L^3$-Net and the spectrogram-based network by a significant margin. Fig.~\ref{fig:qual-eval} presents sample results for the affective correspondence prediction task for each emotion class. The top row shows that the ACP-Net correctly predicts true correspondence between a low tempo song with negative emotion label and negative images. The middle row has a upbeat country song, which has been predicted to have correspondence with \textit{positive} images by the ACP-Net. Note that a true correspondence has been incorrectly predicted for the image with label \textit{anger}. This may be due to fact that the image is not a natural image but a synthetic one. The bottom row has a soothing song for which the ACP-Net did not perform very well. This song has been incorrectly associated with images that are labeled as \emph{awe}, \emph{sadness}, \emph{fear}, \emph{anger}.\\

 \begin{table}[!t]
\setlength{\tabcolsep}{2.2pt}
\caption{Emotion recognition results for individual modalities}
\label{tab:emo_uni}
    \vspace{2mm}
    \centering
    \begin{tabular}{l c}
    \multicolumn{2}{c}{\emph{Emotion in images}}\\
    \hline
    \bf Method & {\bf Accuracy} (in $\%$) \\
    \hline
    Random & 12.5 \\
    Art-based emotion features \cite{zhao2014exploring} & 46.5  \\
    AlexNet \cite{rao2016learning} & 58.6\\
    ACP-Net features + MLP & 40.4 \\
    \hline \vspace{-2mm} \\
    \multicolumn{2}{c}{\emph{Emotion in music}}\\
    \hline
    \bf Method & {\bf Accuracy} (in $\%$) \\
    \hline
    Random & 33.3  \\
    Spectrogram + CNN & 47.8\\
    ACP-Net features + MLP & 63.5\\
    \hline
    \end{tabular}
\end{table}

\vspace{-2mm}\noindent\textbf{Emotion recognition in images and music:} Note that ACP-Net does not use any emotion label explicitly for training. Thus, it is important to investigate if the network is actually able to capture any emotion-specific information. To study this we use the learned embeddings to perform emotion classification in their respective modalities. For images, emotion recognition is performed on the Image Emotion database \cite{HailinJin}. The $1024$-dimensional vector output by the image subnetwork in the ACP-Net is fed to a multilayer perceptron (MLP) to perform an 8-class emotion classification. The eight classes correspond to the original emotion labels in Image Emotion Database (see Table \ref{tab:emo_uni}). 
The performance of the ACP-Net features is compared against several existing models trained particularly for 8-class emotion recognition in images. Results show that the ACP-Net features capture significant information about emotion. Also note that the ACP-Net features were obtained from a network which is designed to discriminated among only 3 classes, but still, ACP-Net features perform well on the 8-class classification. 
\par Similar to the emotion recognition task in images, the $1024$-dimensional vector output by music subnetwork is used to perform a 3-class (\textit{positive}, \textit{neutral}, \textit{negative}) emotion recognition in music. We used an MLP with 2 hidden layers (with 512 and 32 nodes), an input and an output layer. The classification accuracy achieved is $63.56\%$ (see Table \ref{tab:emo_uni}). 
The reasonable performance of the ACP-Net features for emotion recognition tasks indicate that ACP-Net is indeed able to capture emotion-related information while learning to perform a much higher level task. 
\section{Conclusion}
This paper introduced the task of affective correspondence learning between music and image, and proposed a deep architecture (ACP-Net) to accomplish the task. The ACP-Net uses two subnetworks corresponding to image and music to project the data from individual modalities to a common emotion space. Several fusion layers are used to learn if the input image-music pair is similar or not in terms of their emotion content. We also constructed a large scale database containing music and images with emotion labels. Our experiments on this database show that the proposed network can achieve good prediction accuracy on this challenging task, while learning meaningful representation of emotion from individual modalities that can be useful for other related tasks, such as emotion recognition. A potential application of our network would be in crossmodal media retreival, where explicit emotion labels may not be available. Future work could be directed towards incorporating the song transcripts (if available) to capture the emotion content in music better. 
\newpage
\balance
\bibliographystyle{IEEEbib}
\bibliography{strings}

\begin{thebibliography}{10}

\bibitem{joshi2011aesthetics}
Dhiraj Joshi, Ritendra Datta, Elena Fedorovskaya, Quang-Tuan Luong, James~Z
  Wang, Jia Li, and Jiebo Luo,
\newblock ``Aesthetics and emotions in images,''
\newblock {\em IEEE Signal Processing Magazine}, vol. 28, no. 5, pp. 94--115,
  2011.

\bibitem{you2015robust}
Q~You, J~Luo, H~Jin, and J~Yang,
\newblock ``Robust image sentiment analysis using progressively trained and
  domain transferred deep networks.,''
\newblock in {\em AAAI}, 2015, pp. 381--388.

\bibitem{kahou2013combining}
S~E Kahou, C~Pal, X~Bouthillier, P~Froumenty, {\c{C}}~G{\"u}l{\c{c}}ehre,
  R~Memisevic, P~Vincent, A~Courville, Y~Bengio, R~C Ferrari, et~al.,
\newblock ``Combining modality specific deep neural networks for emotion
  recognition in video,''
\newblock in {\em ACM Int conf on multimodal interaction (ICMI)}, 2013, pp.
  543--550.

\bibitem{schuller2003hidden}
B~Schuller, G~Rigoll, and M~Lang,
\newblock ``Hidden markov model-based speech emotion recognition,''
\newblock in {\em Int Conf on Multimedia and Expo (ICME)}. IEEE, 2003, vol.~1,
  pp. I--401.

\bibitem{yang2011ranking}
Y~H Yang and H~H Chen,
\newblock ``Ranking-based emotion recognition for music organization and
  retrieval,''
\newblock {\em IEEE Trans on Audio, Speech, and Language Processing}, vol. 19,
  no. 4, pp. 762--774, 2011.

\bibitem{socher2013zero}
R~Socher, M~Ganjoo, C~D Manning, and A~Ng,
\newblock ``Zero-shot learning through cross-modal transfer,''
\newblock in {\em Advances in neural information processing systems}, 2013, pp.
  935--943.

\bibitem{owens2016ambient}
A~Owens, J~Wu, J~H McDermott, W~T Freeman, and A~Torralba,
\newblock ``Ambient sound provides supervision for visual learning,''
\newblock in {\em European Conference on Computer Vision}, 2016, pp. 801--816.

\bibitem{goyal2016multimodal}
A~Goyal, N~Kumar, T~Guha, and S~S Narayanan,
\newblock ``A multimodal mixture-of-experts model for dynamic emotion
  prediction in movies,''
\newblock in {\em IEEE Int Conf on Acoustics, Speech and Signal Processing
  (ICASSP)}, 2016, pp. 2822--2826.

\bibitem{cao2016deep}
Y~Cao, M~Long, J~Wang, Q~Yang, and P~S Yu,
\newblock ``Deep visual-semantic hashing for cross-modal retrieval,''
\newblock in {\em ACM Int Conf on Knowledge Discovery and Data Mining (KDD)},
  2016, pp. 1445--1454.

\bibitem{hong2017deep}
S~Hong, W~Im, and H~S Yang,
\newblock ``Deep learning for content-based, cross-modal retrieval of videos
  and music,''
\newblock {\em arXiv preprint arXiv:1704.06761}, 2017.

\bibitem{owens2018audio}
A~Owens and A~A Efros,
\newblock ``Audio-visual scene analysis with self-supervised multisensory
  features,''
\newblock {\em arXiv preprint arXiv:1804.03641}, 2018.

\bibitem{nagrani2018seeing}
Arsha Nagrani, Samuel Albanie, and Andrew Zisserman,
\newblock ``Seeing voices and hearing faces: Cross-modal biometric matching,''
\newblock in {\em Proceedings of the IEEE Conference on Computer Vision and
  Pattern Recognition}, 2018, pp. 8427--8436.

\bibitem{arandjelovic2017look}
R~Arandjelovic and A~Zisserman,
\newblock ``Look, listen and learn,''
\newblock in {\em IEEE Int Conf on Computer Vision (ICCV)}, 2017, pp. 609--617.

\bibitem{arandjelovic2017objects}
R~Arandjelovic and A~Zisserman,
\newblock ``Objects that sound,''
\newblock {\em arXiv preprint arXiv:1712.06651}, vol. 3, no. 10, 2017.

\bibitem{HailinJin}
Quanzeng You, Jiebo Luo, Hailin Jin, and Jianchao Yang,
\newblock ``Building a large scale dataset for image emotion recognition: The
  fine print and the benchmark.,''
\newblock in {\em AAAI}, 2016, pp. 308--314.

\bibitem{bertin2011million}
T~Bertin-Mahieux, D~PW Ellis, B~Whitman, and P~Lamere,
\newblock ``The million song dataset,''
\newblock in {\em ISMIR}, 2011, vol.~2, p.~10.

\bibitem{szegedy2016rethinking}
Christian Szegedy, Vincent Vanhoucke, Sergey Ioffe, Jon Shlens, and Zbigniew
  Wojna,
\newblock ``Rethinking the inception architecture for computer vision,''
\newblock in {\em Proceedings of the IEEE conference on computer vision and
  pattern recognition}, 2016, pp. 2818--2826.

\bibitem{russakovsky2015imagenet}
Olga Russakovsky, Jia Deng, Hao Su, Jonathan Krause, Sanjeev Satheesh, Sean Ma,
  Zhiheng Huang, Andrej Karpathy, Aditya Khosla, Michael Bernstein, et~al.,
\newblock ``Imagenet large scale visual recognition challenge,''
\newblock {\em International Journal of Computer Vision}, vol. 115, no. 3, pp.
  211--252, 2015.

\bibitem{west2004features}
Kristopher West and Stephen Cox,
\newblock ``Features and classifiers for the automatic classification of
  musical audio signals.,''
\newblock in {\em ISMIR}, 2004.

\bibitem{lu2006automatic}
Lie Lu, Dan Liu, and Hong-Jiang Zhang,
\newblock ``Automatic mood detection and tracking of music audio signals,''
\newblock {\em IEEE Transactions on audio, speech, and language processing},
  vol. 14, no. 1, pp. 5--18, 2006.

\bibitem{scherer1996adding}
Klaus~R Scherer,
\newblock ``Adding the affective dimension: a new look in speech analysis and
  synthesis.,''
\newblock in {\em ICSLP}, 1996.

\bibitem{jiang2002music}
Dan-Ning Jiang, Lie Lu, Hong-Jiang Zhang, Jian-Hua Tao, and Lian-Hong Cai,
\newblock ``Music type classification by spectral contrast feature,''
\newblock in {\em Multimedia and Expo, 2002. ICME'02. Proceedings. 2002 IEEE
  International Conference on}. IEEE, 2002, vol.~1, pp. 113--116.

\bibitem{harte2006detecting}
Christopher Harte, Mark Sandler, and Martin Gasser,
\newblock ``Detecting harmonic change in musical audio,''
\newblock in {\em Proceedings of the 1st ACM workshop on Audio and music
  computing multimedia}. ACM, 2006, pp. 21--26.

\bibitem{kingma2014adam}
Diederik~P Kingma and Jimmy Ba,
\newblock ``Adam: A method for stochastic optimization,''
\newblock {\em arXiv preprint arXiv:1412.6980}, 2014.

\bibitem{gupta2014multimodal}
R~Gupta, N~Malandrakis, B~Xiao, T~Guha, M~Van~Segbroeck, M~P Black,
  A~Potamianos, and S~Narayanan,
\newblock ``Multimodal prediction of affective dimensions and depression in
  human-computer interactions,''
\newblock in {\em AVEC@ACM MM}. ACM, 2014, pp. 33--40.

\bibitem{zhao2014exploring}
Sicheng Zhao, Yue Gao, Xiaolei Jiang, Hongxun Yao, Tat-Seng Chua, and Xiaoshuai
  Sun,
\newblock ``Exploring principles-of-art features for image emotion
  recognition,''
\newblock in {\em Proceedings of the 22nd ACM international conference on
  Multimedia}. ACM, 2014, pp. 47--56.

\bibitem{rao2016learning}
Tianrong Rao, Min Xu, and Dong Xu,
\newblock ``Learning multi-level deep representations for image emotion
  classification,''
\newblock {\em arXiv preprint arXiv:1611.07145}, 2016.

\end{thebibliography}
\end{document}